\documentstyle[12pt]{article}
  \advance\oddsidemargin  by -1in
  \advance\evensidemargin by -1in
\def\lsim{\mathrel{\rlap{\lower4pt\hbox{\hskip1pt$\sim$}}
    \raise1pt\hbox{$<$}}}         
\def\gsim{\mathrel{\rlap{\lower4pt\hbox{\hskip1pt$\sim$}}
    \raise1pt\hbox{$>$}}}         

\def\Pom{ I\!\!P  }

\def\be{\begin{equation}}
\def\ee{\end{equation}}
\def\bq{\begin{eqnarray}}
\def\eq{\end{eqnarray}}
\def\bm{\boldmath}
\def\bvec{\mbox{\bm $b$}}

\begin{document}

\pagestyle{empty}

\begin{flushright}
DFTT 28/97 \\
May 1997 \\
\end{flushright}

\vspace{2cm}

\begin{center}

{\LARGE  \bf On the  low-$\mbox{\bm $x$}$
behavior \\ of nuclear shadowing \\}

\vspace{2 cm}

V.~BARONE$^{a,}$\footnote{Also at II Facolt{\`a} di Scienze MFN, 15100
Alessandria, Italy.}, M.~GENOVESE$^{b,}$\footnote{Supported by 
EU Contract ERBFMBICT950427.}
\vskip 0.2cm

{\it $^{a}$Dipartimento di
Fisica Teorica dell' Universit\`a \\
and INFN, Sezione di
Torino,   I--10125 Turin, Italy \medskip\\

$^{b}$ Institut des Sciences Nucl\'eaires \\
Universit\'e Joseph Fourier--IN2P3-CNRS \\
53, avenue des Martyrs, F-38026 Grenoble Cedex,  
France\\}

\vskip 3cm

{\bf Abstract}

\end{center}

We calculate the $x$ dependence of nuclear shadowing 
at moderate values of $Q^2$ by 
using HERA diffractive data. 
We show that 
no decrease 
of shadowing occurs down to  
very low $x$  ($x \simeq 10^{-4}$). 

\newpage
\pagestyle{plain}
\baselineskip 20pt

The possibility of studying nuclear shadowing
({\it i.e.} the depletion 
of bound nucleon structure functions ($F_2^A$) 
with respect to 
the free nucleon ones ($F_2^N$))
at HERA \cite{HERA} has prompted new interest on this subject. 
Different models and interpretations have been suggested in the 
past years to explain 
nuclear shadowing \cite{nos}.
A nuclear HERA program would allow
extending the experimental investigation
to very low $x$, thus offering the chance
of a 
deeper understanding of the 
phenomenon. 

One recent interesting prediction \cite{Boris} 
is that, at low $x$ and for moderately large $Q^2$ 
($Q^2 \sim 10$ GeV$^2$)
 shadowing 
stops its rise,  
starts decreasing and eventually vanishes. 
This effect arises from the different low-$x$ behavior of the 
inclusive and the diffractive structure functions. 
In fact, at  
$x \ll 1$, whereas 
$F_2(x,Q^2) \sim (1/x)^{\Delta(Q^2)}$, with 
a `hard' intercept $\Delta(Q^2) \approx 
0.3-0.4$, the shadowing term $\Delta F_2^{\rm sh} 
\equiv F_2^A - F_2^N$, 
which is proportional to the diffractive 
structure function, is dominated by soft physics and
has an asymptotic $x$ dependence  of the form $\Delta F_2^{\rm sh} 
\sim
 (1/x)^{2 \Delta(\mu^2)}$,
with $\Delta(\mu^2) \approx 0.1$, where $\mu^2$ is the 
typical scale of soft 
interactions.

While the argument above is 
 rather general and correct, the specific 
finding of Ref.~\cite{Boris}, {\it i.e.}
that at $Q^2 \sim 10$ GeV$^2$
shadowing reaches its
largest value already at $x \approx 10^{-2}$ 
and then begins to decrease towards 
smaller $x$, is based 
on some restrictive assumptions.  
We shall call $x_0$ the $x$ value where shadowing attains 
its maximum. 
The value of $x_0$ is found in \cite{Boris} to be 
almost independent of $Q^2$ (for $Q^2 \gsim
5$ GeV$^2$) and of the atomic mass.  
The prediction of \cite{Boris} is that 
in the region between $x = 10^{-4}$ and $x=10^{-2}$ shadowing 
should be a decreasing function of $x$ towards 
$x \rightarrow 0$. It is evident that such a behavior
should easily be observable at HERA.

The results of Ref.~\cite{Boris} rely on two
assumptions: 

1) the shadowing term,  which is related to the 
diffractive structure function, 
 is given by a simple power--like
parameterization 
\be
\Delta F_2^{\rm sh} \sim  x^{2 - 2 \alpha_{\Pom}(0)}
\label{1}
\ee
with $\alpha_{\Pom}(0) \simeq 1.1$;

2) the mass $M^2$ of the diffractively 
produced states is kept fixed and 
equal to $Q^2$. 

Both these assumptions are questionable.  
The first one is true only asymptotically, as  $x \rightarrow 0$,  
and cannot be used to draw any conclusion 
about the $x$-region between $10^{-4}$ and $10^{-2}$. 
The second assumption 
is not valid when the dynamics is 
dominated by the triple pomeron contribution, 
which provides a nonvanishing high--mass tail to the 
distribution of the diffracted states. 

The purpose of this letter is to carry out 
a more precise phenomenological analysis
of nuclear shadowing, by relying on 
the diffractive structure function measurements 
of HERA. Our main concern will be with the 
$x$--dependence of shadowing, rather than with its 
absolute normalization, which is at present unpredictable
due to experimental and theoretical uncertainties. 
Our 
calculation, which does not make use of the assumptions 
of Ref.~\cite{Boris},  leads 
to the conclusion that the onset of the decrease of shadowing
is likely to be around or smaller than $x_0 =  10^{-4}$ and 
thus hardly reachable at HERA. In other terms, we predict that 
no decrease of shadowing will be observed at HERA. 

Let us start from the well-known relation between nuclear shadowing and 
diffraction dissociation established long ago by Gribov 
\cite{Gribov,KK}.  
In virtual-photon--nucleus scattering
the nuclear cross section is given by \cite{nos,nos2}
\be
\sigma^{\gamma^{*}A} = A \, \sigma^{\gamma^{*}N} - 
4 \pi \, \frac{A-1}{A} \, \int {\rm d} M^2 \left. 
\frac{{\rm d}^2 \sigma^{D}}{{\rm d}M^2 \, {\rm d}t}\right \vert_{t=0} 
\, \int {\rm d}^2 \bvec \, \vert \Phi(k_L, \bvec) \vert^2 \,  
+ \ldots \,,
\label{3}
\ee
where ${\rm d}^2 \sigma^D/dM^2 dt$ 
is the $\gamma^{*}N$ diffraction dissociation 
cross section ($M^2$ being the invariant mass of the excited hadronic 
states), and the longitudinal form factor $\Phi$, which is 
function of the impact parameter $\bvec$ and of the longitudinal 
momentum of the recoil proton $k_L = x m_N (1 + M^2/Q^2)$, 
is related to the nuclear density $\rho_A$ by 
\be
\Phi(k_L, \bvec) = \int {\rm d} z \, \rho_A(\bvec, z) \, 
{\rm e}^{i k_L z}\,.
\label{4}
\ee

The dots in eq.~(\ref{3}) represent the higher rescattering terms, 
which are non negligeable for heavy nuclei. The simplest 
way to 
take them approximately into account is \cite{Kwi} to 
introduce an eikonal factor 
${\rm e}^{- \sigma_{\rm eff} T(\bvec)/2}$ in the integration over the 
impact parameter in (\ref{3}). $T(\bvec)$ is the nuclear thickness
\be 
T(\bvec) = \int_{-\infty}^{+\infty} {\rm d}z \, \rho_A(\bvec, z)\,;
\label{4.0.1}
\ee 
$\sigma_{\rm eff}$ is the effective cross section for the 
interaction of the diffracted states with the nucleon, given by 
\be
\sigma_{\rm eff} = 16 \pi \frac{1}{\sigma^{\gamma^{*}N}}\, \left.
\frac{{\rm d}\sigma^{D}}{{\rm d}t} \right \vert_{t=0}\,.
\label{4.1}
\ee
At small $x$ this cross section  turns out to be almost 
$x$ independent.
In 
practical calculations it will be taken as a constant
(see below). 

In terms of the inclusive structure functions per nucleon 
$F_2^{N,A}$, and of the diffractive structure function  
\be
F_2^{D(4)}(\beta, Q^2, x_{\Pom}, t=0) = 
\frac{4 \pi^2 \alpha_{\rm em}}{Q^2} \, \left.
\frac{{\rm d}^4 \sigma^D}{{\rm d}\beta \, {\rm d} Q^2 
\, {\rm d}x_{\Pom} \, {\rm d}t} \right \vert_{t=0}\,,
\label{4.2}
\ee
where ($W^2$ is the squared center-of-mass energy of the $\gamma^{*} N$
system)  
\be
x_{\Pom} \equiv \frac{M^2 + Q^2}{W^2 + Q^2}, \,\,\,\,
\beta \equiv  \frac{x}{x_{\Pom}}\,,
\label{4.3}
\ee
eq.~(\ref{3}) becomes ($k_L = m_N x_{\Pom}$)
\bq
F_2^{A}(x, Q^2) &=& F_2^{N} (x, Q^2) \nonumber \\
&-& 4 \pi  \frac{A-1}{A^2}  \,
\int {\rm d} x_{\Pom} \, F_2^{D(4)}(\beta, Q^2, x_{\Pom}, t=0)
\, \int {\rm d}^2 \bvec 
\, {\rm e}^{- \frac{\sigma_{\rm eff}}{2} T(\bvec)} \, 
\vert \Phi(x_{\Pom}, \bvec) \vert^2 \, .
\label{4.4}
\eq

In principle,  
the diffractive structure function $F_2^{D(4)}$ can be obtained from the 
experiment. However, what the present experimental analyses provide 
is only the diffractive structure function integrated over $t$
\be
F_2^{D(3)} (\beta, Q^2, x_{\Pom}) = 
\int_{0}^{\vert t \vert_{\rm max}} {\rm d} \vert t \vert \, 
F_2^{D(4)}(\beta, Q^2, x_{\Pom}, t)\,,
\label{4.5}
\ee
with $t_{\rm max} \simeq  0.5$ GeV$^2$. 

The 
ZEUS parametrization \cite{ZEUSFD} for 
$F_2^{D(3)}$ has the form
\bq
F_2^{D (3)} &=& F_v^{D(3)} + F_s^{D(3)} \nonumber \\ 
&=&   A \, x_{\Pom}^{-a} \, [ \beta (1- \beta) + 
\frac{C}{2} \,  (1- \beta)^2 ] \, 
\label{5} 
\eq
where 
$a=1.30 \pm 0.08 ^{+ 0.08} _{ - 0.14}$, 
$A = 0.018 \pm 0.001 \pm 0.005$,
$C= 0.57 \pm 0.12 \pm 0.22$. 
The exponent $a$ is found to be 
essentially independent of $\beta$. 
A more recent preliminary analysis \cite{Chi} gives a smaller value 
for $a$: $a \sim 1.1-1.2$ in the $Q^2$ range $10-20$ GeV$^2$. 
In eq.~(\ref{5}) we separated a `valence' part 
$F_v^{D(3)} \propto \beta (1- \beta)$ and a `sea' part
$F_s^{D(3)} \propto (1- \beta)^2$. In the language of the color dipole 
model \cite{NZ}, the valence corresponds to the 
lowest Fock state ($q \bar q$)
of the virtual photon, whereas the sea corresponds to higher 
Fock states ($q \bar q g ...$), which represent the
triple pomeron contribution of Regge theory. 
Notice that 
factorization breaking effects, predicted 
in Ref. \cite{GNZ}, which would imply a non--universal flux 
factor for the valence and sea components,  
are not yet observable. 

In order to derive $F_2^{D(4)}(\beta, Q^2, x_{\Pom}, t=0)$
from the measured $F_2^{D(3)}(\beta, Q^2, x_{\Pom})$, eq.~(\ref{5}), 
we assume a simple peripheral $t$-dependence of the form 
\be
F_{v,s}^{D(4)}(\beta, Q^2, x_{\Pom}, t) = 
F_{v,s}^{D(4)}(\beta, Q^2, x_{\Pom}, t=0) \, 
{\rm e}^{- (B_{v,s} + 2 \alpha'_{\Pom} \log{\frac{1}{x_{\Pom}}} ) 
\, \vert t \vert } \,, \,
\label{10}
\ee
where $\alpha'_{\Pom} \simeq 0.5$ and the slopes $B_{v,s}$
are borrowed from hadron scattering and real photoproduction. 
We use $B_s \simeq 6$ GeV$^{-2}$,  
$B_v \simeq 12$ GeV$^{-2}$.  

In eq.~(\ref{4.4}) 
 $F_2^{D (3)}(\beta,x_{\Pom},Q^2)$ is integrated
over $x_{\Pom}$ between $x$ and 1. However, 
due to the selection of the rapidity gap events, 
there is an 
experimental upper cutoff on $x_{\Pom}$: $x_{\Pom}^{\rm c} 
= 0.04$. We checked that in the $x$ region of interest the 
large--$x_{\Pom}$ tail neglected in the integration is irrelevant. 

For consistency,  
we used the ZEUS parametrization
for $F_2^{N}(x,Q^2)$ \cite{ZEUSF}.
Our predictions do not depend 
on the parametrization adopted
for $F_2$ (the same results are obtained using MRS(R1) \cite{MRS}).
As for the nuclear part of the calculation, we used a 
Fermi--type nuclear density.

The two main sources of uncertainty in our calculation are: 
{\it i)} the effective cross section $\sigma_{\rm eff}$ 
(a constant value, specified below, is used); {\it ii)} the 
large experimental error on $a$, the exponent of the 
so-called pomeron flux, see eq.~(\ref{5}).
As we shall see, these uncertainties prevent us from 
predicting the absolute amount of shadowing, although they do not 
affect the qualitative features of the $x$--dependence of 
shadowing, that we are most interested in. Both the normalization
and the $x$--dependence of shadowing depend very little on the 
other parameters appearing in the calculation. In particular we checked
that even a large variation of the slopes $B_{v,s}$ changes by 
no more than few percent the predicted shadowing.

Let us come now to the results.

In Fig.~1 we plot the ratio $F_2^A/F_2^N$ 
for calcium. 
The diffractive structure function used in the calculation 
is 
given by the ZEUS parametrization 
(\ref{5}), which is valid at moderately large 
$Q^2$ ($Q^2 \simeq 10-30$ GeV$^2$). We use
the central values of the ZEUS fit
for the coefficients 
$A$ and $C$, quoted after eq.~(\ref{5}). We allow 
the exponent of the pomeron flux $a$ 
to vary around a value ($a=1.2$) which is 
smaller than the central value ($a=1.3$) of the ZEUS analysis 
\cite{ZEUSFD} and closer to the most recent finding 
($a \simeq 1.1-1.2$) \cite{Chi}. For comparison, 
the value used in Ref.~\cite{Boris} is 1.20. 
For the effective cross 
section we take $\sigma_{\rm eff} =10$ mb. 
The $x$ range shown in figure is the one allowed by the present 
experimental fits on the diffractive structure function. 
No decrease of shadowing occurs down to 
$x \simeq 10^{-4}$. 

In Fig.~2 we set $a=1.2$ and we vary the coefficient $C$ 
of the sea component of the diffractive structure function 
(\ref{5}) within the errors of the ZEUS fit \cite{ZEUSFD}. 
We take $\sigma_{\rm eff} =10$ mb. 
Again, the shadowing curve is at most (when 
$C$ is small) rather flat 
towards $x=10^{-4}$ but does not exhibit any sensible decrease. 

Our main  finding is therefore that the onset of shadowing 
saturation is at  
much lower $x$ than argued in Ref.~\cite{Boris} 
($x_0 \lsim  10^{-4}$) and is 
outside  the range of investigation of HERA experiments.
Only a combination of unlikely circumstances 
($a$ and $C$ very small, $a \lsim 1.1$, $C \simeq 0$, that 
is a pomeron flux less singular than the Donnachie--Landshoff
one and an almost vanishing sea component in the 
diffractive structure function)  would produce a 
visible decrease of nuclear shadowing in the region above 
$x  = 10^{-4}$.

In Fig.~3 we illustrate the dependence of 
our results 
on the effective cross section $\sigma_{\rm eff}$, for 
two nuclei (carbon and calcium). 
The shadowing curves with 
two choices of 
$\sigma_{\rm eff}$ (12 and 15 mb) 
and $a=1.2$ in eq.~(\ref{5}) 
are shown. 
We see that the value of $\sigma_{\rm eff}$ 
affects the absolute normalization of nuclear 
shadowing but    
not its $x$--dependence.
For light nuclei, such as carbon, 
not even the normalization depends on $\sigma_{\rm eff}$. 
Thus the theoretical uncertainty related to the choice 
of $\sigma_{\rm eff}$ does not spoil our conclusions 
about the $x$ behavior of the shadowing curve.


Considering the weak dependence of the shape of 
shadowing on the theoretical ingredients 
of the present calculation 
(the effective cross section for multiple rescattering, 
the hadronic slopes and the longitudinal form factor), 
we can say that our prediction is  dictated essentially by
the experimental measurements of $F_2^D$ and therefore 
is a model independent result. Moreover, the qualitative behavior 
of shadowing that we found is stable against a large variation of the 
exponent of the pomeron flux and of the size of the 
sea diffractive structure function. Hence, any more 
precise
determination of the diffractive structure function 
should not change our main finding. 

Finally, we just mention that results qualitatively 
similar to those reported above can be obtained by using
$F_2^{D (3)}$
evaluated 
in BFKL--type models, such as the one of Ref.~\cite{GNZ}:  
only the absolute size of shadowing turns out to be larger.

In conclusion, let us summarize our results. 
We 
studied the low-$x$ behavior of nuclear shadowing at moderately 
large $Q^2$. We carried out a model independent calculation 
by using the experimental data on the diffractive structure function
and the well established relation between diffraction and shadowing. 
No restrictive assumptions 
were made. Our conclusion is that, to the best of the 
present experimental knowledge, no decrease of shadowing 
occurs above $x \simeq 10^{-4}$, that is 
in the kinematic region accessible at HERA. Obviously, if the 
present experimental results on the different low-$x$ behavior 
of $F_2^N$ and $F_2^{D(3)}$ will be confirmed,  a decrease of 
shadowing does take place, but only at 
very small $x$, beyond the HERA range.

\vspace{1cm}
We thank M.~Arneodo, E.~Barberis, A.~Solano
and A.~Staiano for many useful 
discussions on the ZEUS analyses. 

\pagebreak

\pagebreak

\begin{center}

{\large \bf Figure captions}

\end{center}

\vspace{1cm}

\begin{itemize}

\item[Fig.~1]
The ratio $F_2^A(x)/F_2^N(x)$ for calcium
in the $Q^2$ range 10-30 GeV$^2$. The three curves correspond 
to three different values of the exponent of the pomeron 
flux in the ZEUS fit of the diffractive structure function
eq.~(\ref{5}): $a=1.15$ (dotted line), $a=1.20$ (dashed line), 
$a=1.25$ (solid line). The other parameters
used are specified in the text. 

\item[Fig.~2]
The ratio $F_2^A(x)/F_2^N(x)$ for calcium
in the $Q^2$ range 10-30 GeV$^2$. The solid 
curve is obtained using $C = 0.57$ in 
the ZEUS fit, see eq.~(\ref{5}). The dashed
curves correspond to $C =0.57 \pm 0.25$ (errors 
added in quadrature).  
The other parameters are specified in the text.

\item[Fig.~3]
The ratio $F_2^A(x)/F_2^N(x)$ for carbon 
(upper pair of curves) and calcium (lower pair 
of curves) 
in the $Q^2$ range 10-30 GeV$^2$. 
The solid and dashed lines correspond to 
$\sigma_{\rm eff} = 10$ mb and $\sigma_{\rm eff} =15$ mb, 
respectively.

\end{itemize}

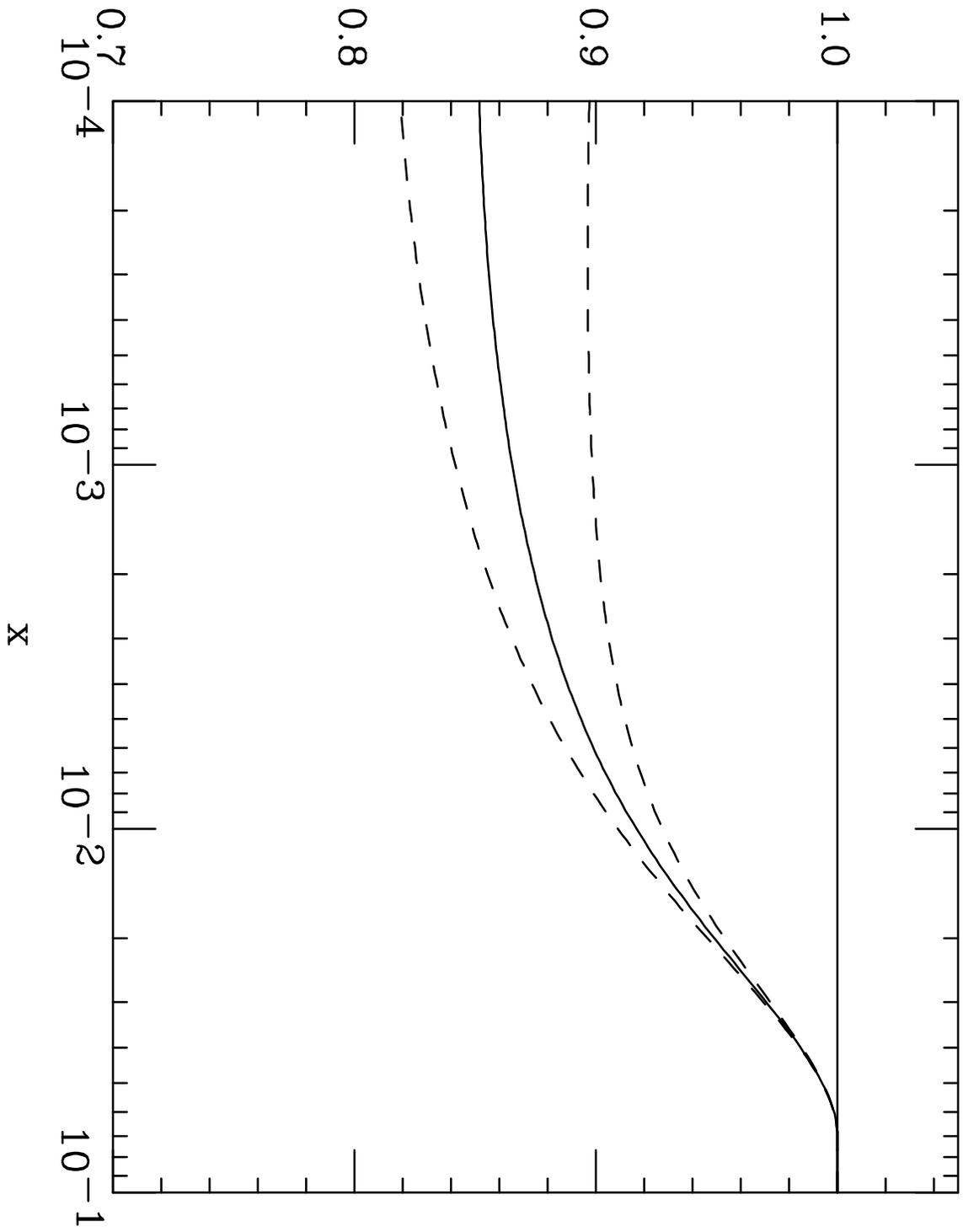
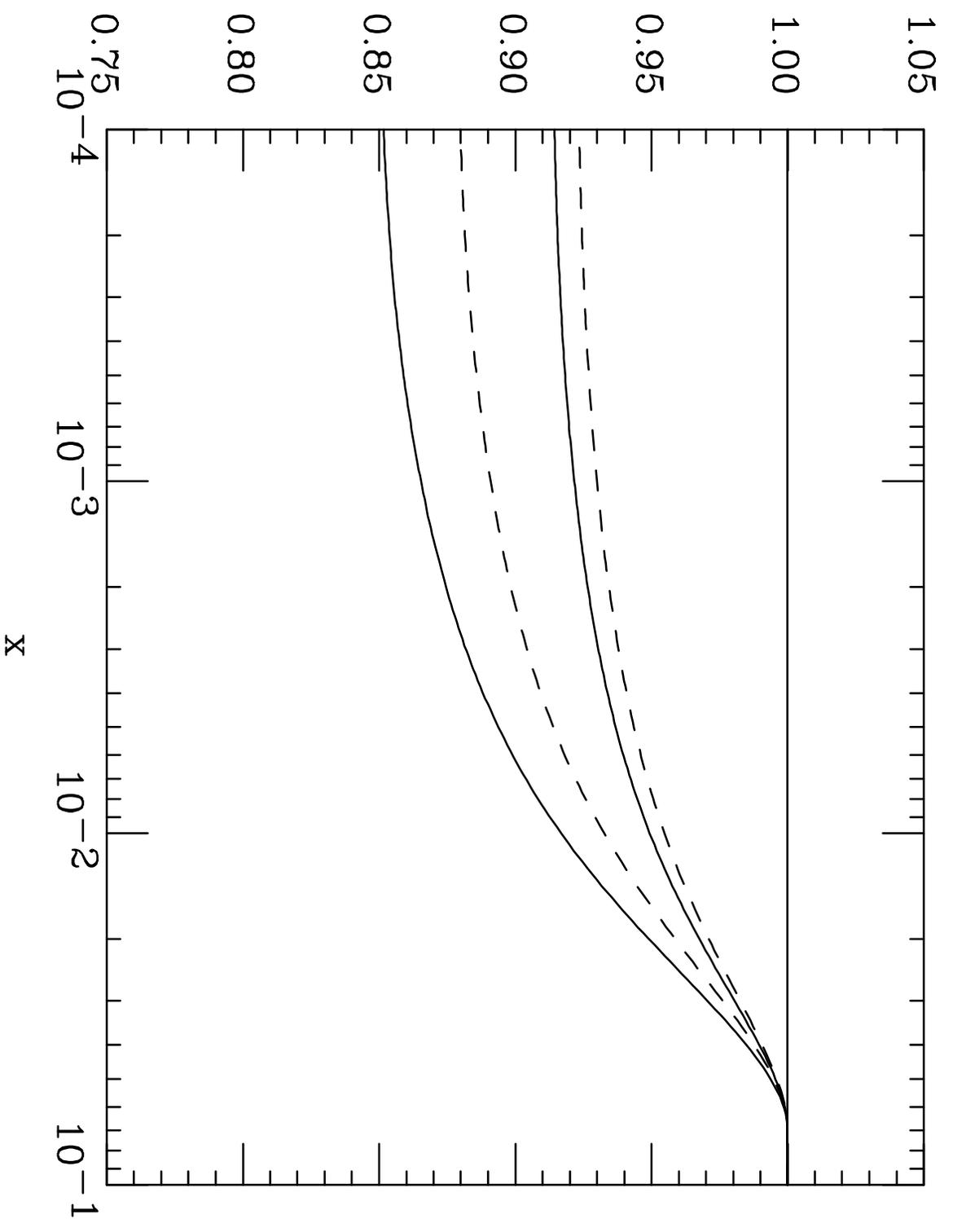
\end{document}